\documentclass{AIMS}
\usepackage{amsmath}
  \usepackage{paralist}
  \usepackage{graphics} 
  \usepackage{epsfig} 
\usepackage{tikz}  
\usepackage{graphicx}  
\usepackage{epstopdf}
\usepackage[utf8]{inputenc}
\usepackage{authblk}
\usepackage{array}

 \usepackage[colorlinks=true]{hyperref}
\hypersetup{urlcolor=blue, citecolor=red}

    \def\ppages{}
     \def\DOI{}
\newcolumntype{K}[1]{>{\centering\arraybackslash}p{#1}}

\theoremstyle{definition}

\title[Estimation of Dengue reproduction number] 
      {Estimation of reproduction number and non stationary spectral analysis of Dengue epidemic}

\author[first-name1 last-name1 and first-name2 last-name2]{}


 \email{endurimuralikrishna@iitgn.ac.in}
 \email{shiva.jolad@iitgn.ac.in}

\thanks{ We gratefully acknowledge the partial funding for this work from SERB-DST India through project number SB/FTP/PS-033/2013}

\thanks{$^*$ Corresponding author: Shivakumar Jolad}

\begin{document}
\maketitle

\centerline{\scshape Murali Krishna Enduri}
\medskip
{\footnotesize
 \centerline{Indian Institute of Technology Gandhinagar}
   \centerline{ Gandhinagar, Gujarat- 382355, INDIA}
} 

\medskip

\centerline{\scshape Shivakumar Jolad $^*$}
\medskip
{\footnotesize

\centerline{Indian Institute of Technology Gandhinagar}
   \centerline{ Gandhinagar, Gujarat- 382355, INDIA}
}

\bigskip


\begin{abstract}
In this work we analyze the post monsoon Dengue outbreaks  by analyzing the transient and long term dynamics of Dengue incidences and its environmental correlates in Ahmedabad city in western India from 2005-2012. We calculate the reproduction number $R_p$ using the growth rate of post monsoon Dengue outbreaks and biological parameters like host and vector incubation periods and vector mortality rate, and its uncertainties are estimated through Monte-Carlo simulations by sampling parameters from their respective probability distributions. Reduction in Female \textit{Aedes} mosquito density required for an effective prevention of Dengue outbreaks is also calculated. The non stationary pattern of Dengue incidences and its climatic correlates like rainfall temperature is analyzed through Wavelet based methods.  We find that the mean time lag between between peak of monsoon and Dengue is 9 weeks. Monsoon and Dengue cases are phase locked from 2008-2012 in the 16-32 weeks band. 
The duration of post monsoon outbreak has been increasing every year, especially post 2008, even though the intensity and duration of monsoon has been decreasing. Temperature and Dengue incidences show correlations in the same band, but phase lock is not stationary.    
\end{abstract}

\section{Introduction}
Dengue is a vector borne disease endemic in tropical and sub tropical countries worldwide. 
According to WHO estimates, over 2.5 billion people  at the risk  of Dengue  \cite{special2009dengue}, and its repeated outbreaks in recent years has 
caused  a major public health concern in tropical countries. It is  spread by \emph{Aedes Aegepti} and \emph{Aedes Albopictus} mosquitoes carrying 
Dengue virus when they bite humans. There are three main types of Dengue: Dengue Fever (DF), Dengue Hemorrhagic Fever (DHF) and the Dengue Shock 
Syndrome (DSS), caused by four serotypes of Dengue virus DENV1-4 \cite{Nishiura2006, rao2013dynamic}. Out of these DHF and DSS can be fatal.   
At present there is no effective vaccination or treatment for dengue. It is believed that any future dengue vaccination is 
imperfect, \cite{bhamarapravati2000live}  and may not offer protection against all serotypes. Many Dengue infections may not produce severe symptoms, 
there by evading early detection. At present, the only known effective way to prevent dengue outbreak is to devise  vector control strategies and 
minimize vector-human transmission.  In India, Dengue epidemic has spread to almost all the states and is posing a serious public health problems. 
In 2010 alone, 8000 confirmed cases were reported. Being primarily urban epidemic, a sound understanding of the dynamics of the Dengue can help 
in devising strategies for containing the spread urban populations.  In this work, we study the Dengue (of a single Serotype) spread  in a densely populated Ahmedabad city in north western state of Gujarat, India, from 2005-2012.

Many biological, environmental,  and human factors affect the spread of Dengue in urban areas, and the pace, scale and intensity of spread 
varies across different regions \cite{MacieldeFreitas2011452,kuno1995review,Adams2010}. A sound understanding of the spatio-temporal spread of the Dengue can help in devising strategies for containing the spread urban populations.   Several mathematical  models have been  proposed (see \cite{Nishiura2006,derouich2006dengue,Andraud2012}  for reviews) for studying the Dengue. Many of these are the compartmental ordinary differential equation (ODE) models, which  divide the human population into Susceptible, Exposed, Infected, 
and Recovered (SEIR) groups; and vectors into Susceptible, Exposed, Infected  (SEI) groups, and studying their temporal dynamics.  Compartmental models can be used to calculate the 
\textit{basic} reproduction number $R_0$ which gives the number of secondary infections produced by a single infected individual. The $R_0$ is a measure of the severity of the outbreak and  depends upon  biological parameters of vectors and  hosts, the transmission parameters such as biting rate and transmission probability. Some of these parameters may 
 not easily be measured, but are easily captured by the initial growth rate $r$ of Dengue cases \cite{chowell2007estimation, favier2006early}. 
 We make use of $r$ and biological parameters such as incubation period , death rate of mosquitoes and recovery rate of humans to estimate the 
 reproduction number $R_p$ (which excludes people who are immune to disease) for each year. We find that $R_p$ varies  from 1.292($\pm 2$) in 2007 to 1.753($\pm 5$) in 2005. The uncertainties in $R_p$ were estimated using parameter uncertainties (from their respective distributions) and Monte-Carlo simulations. Based on these results we also estimated transmission probability for major outbreak of each year. For an effective control of Dengue outbreaks in Ahmedabad, we estimate that Female Aedes mosquito density should be atleast 35\% below their current level. 

Time series analysis of Dengue cases and its correlates provides crucial information about the complex dynamics of Dengue spread and causal relation with associated climatic variables.  The potentiality of carrier mosquito, \emph{Aedes Aegypti},  to spread Dengue is strongly connected to local weather conditions. Temperature and rain fall plays key role in its life cycle  such as its breeding, biting frequency and extrinsic incubation period \cite{halstead2008dengue}.  Spectral analysis methods such as cross correlations and Fourier analysis has been used to study analyze correlates of  epidemiological and environmental variables, for example to understand aggravation of asthma symptoms and daily minimum temperature \cite{bishop1977statistical}, air pollution and mortality \cite{dominici2002use},  and large-scale climatic oscillations and cholera epidemic \cite{pascual2002cholera}. In this work, we analyzed the cross correlation of Dengue with rainfall and temperature, and found that there is a lag of 9 weeks (20 weeks) between the peak of rainfall (average temperature) and peak of the Dengue incidences.

However, the seasonality and non linearity of the climatic variables and complex nature of human settlement and mobility patterns makes Dengue dynamics non stationary and noisy \cite{cazelles2005nonstationary}, for which the traditional spectral methods are not insightful. Wavelet based techniques are ideally suited for non stationary signals (time series), as they can capture both the  time and periodicity in a single domain \cite{mallat1999wavelet}. Wavelet based methods have recently been used  extensively in epidemiological studies \cite{ grenfell2001travelling,  johansson2009multiyear, broutin2005large, laneri2010forcing} including Dengue and its association with El-nino oscillation\cite{cazelles2005nonstationary} (see \cite{cazelles2007time} for review).  
Through Wavelet spectral analysis, we analyzed the duration of post monsoon Dengue incidences in Ahmedabad and found significant temporal variations every year. Earlier years, the duration of outbreak was short, but later increased to 3-4 months from 2010 onwards.
In contrast, the intensity and duration of monsoon has been decreasing over time.
Wavelet coherence analysis of Dengue incidences and rainfall revealed a 16-32 weeks band between 2008-2012 with high coherence and phase locking. A similar analysis for temperature and Dengue incidences reveals a phase match and then a minor lead of Dengue cases over temperature between 2008-2012 in the 16-32 week band. However, there is a need to look beyond the environmental factors such as Virus evolution and adaptation, socioeconomic factors like urbanization and human mobility \cite{murray2013epidemiology} to understand the hyper-epidemicity of Dengue in Ahmedabad in the last decade.

Our paper is organized as follows: Section I is the current introduction, in Sec.2 we discuss the data set for the study and its spatio-temporal representation. We review the main the compartmental model for Dengue, and  methods for computing  reproduction number using the initial growth rate and disease parameters in Sec. 3. Based on these, our estimates of reproduction numbers and their uncertainties for outbreaks in 
Ahmedabad from 2005-2012 are calculated.  In section 4, we move on to examine the time series characteristics of Dengue and its correlates. A brief review of wavelet methods like continuous wavelet transform, cross wavelet and wavelet coherence is also given.  Corresponding results of time and periodicities of Dengue incidences, cross coherence with rainfall and temperature is examined in detail. We end with summary and main conclusions in Sec.5.

\section{Data}
In this work, we use the data on Dengue incidences, rainfall and temperature data of Ahmedabad from 2005-2012. Ahmedabad is a metro 
city situated on the banks of river Sabarmati in  the north western state of Gujarat. It has a population of about 6 million, with hot 
and semi arid climate, receiving moderate rainfall during monsoon season.  The average summer maximum temperature is $40^0C$.

\begin{figure}[ht!]
\centering 
\fbox{\includegraphics[clip, trim=2cm 15cm 3cm 2.5cm, scale=0.75]{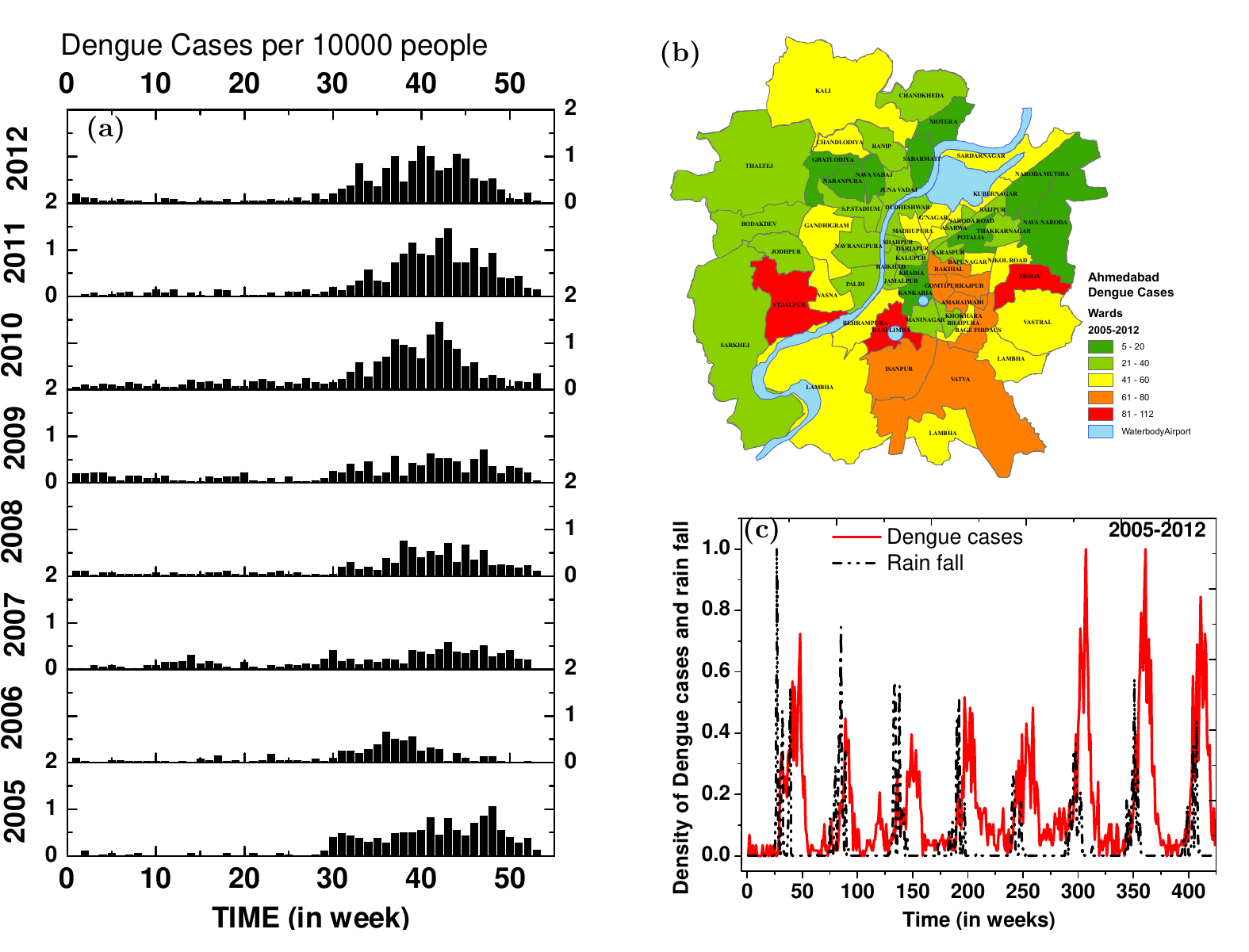}}
\caption{(a) Weekly Dengue cases per 10000 people for Ahmedhabad from 2005-2012.  (b) Dengue cases in Ahmedabad from 2005-2012. (data from  Ahmedabad Municipal Corporation) (c) Comparison of Rain fall, Dengue cases From 2005-2012.}
\label{fig_DGujaratCasesweeklyper10000}
\end{figure}

In Fig.\ref{fig_DGujaratCasesweeklyper10000} (a), Dengue cases (per 10000 population) in Ahmedabad city, reported weekly, for each year from 2005-2012 is shown.  The Dengue cases vary considerably over every year, with primary epidemic outbreak happening in July when the monsoon seasons begins. In Fig.\ref{fig_DGujaratCasesweeklyper10000} (b), we show the Dengue cases at ward level in Ahmedabad from 2005-2012. The densely populated regions in south eastern region have typically higher incidences of Dengue.  Detailed spatial analysis of the dengue spread is beyond the scope of this work.  It reaches peak within 2.5-3 months from the peak of the monsoon season (Ref. Fig \ref{fig_DGujaratCasesweeklyper10000} (c)). In the years  2010-2012, there was marked increase in annual Dengue epidemic cases. In the time series section, we give detailed analysis of the temporal spread of Dengue. We caution the reader that the estimates were based on the officially reported number of cases. Since most Dengue cases are not reported due to mild symptoms and poor record keeping in hospitals in India \cite{Shepard2014,kakkar2012dengue}, the size of epidemic is typically an underestimate.  However, for our calculations of growth rate and spectral analysis, the patterns of change is more important than the scale of change and we assuming that the available data captures the trend in Dengue cases effectively.

\section{Compartmental Model and Reproduction number}

To study the disease dynamics (for a \textit{single serotype}) , it is useful to divide the population into groups according to the state of the disease they are in.
The standard compartmental model for vector borne disease involves dividing humans into Susceptible, Exposed, Infected and Recovered 
(SEIR) groups, assuming homogeneous mixing and vectors into Susceptible, Exposed and Infected (SEI) groups, and study the flow dynamical 
flow between these compartments. Infection dynamics is bidirectional- only infected vectors $I_v$ and infected humans  $I_h$ can transmit 
the dengue virus to susceptible ($S_h,S_v$) population. It depends upon the host parameters: human and vector incubation periods 
$(\tau_e,\tau_i)$, human recovery rate from infection $\gamma_h$, vector morality rate $\mu_v$, and transmission parameters:  mosquito biting rate $C$, The number of female mosquitoes per person $m$, transmissions probabilities $\beta_{hv} (\beta_{hv})$ between vector (human) to human 
(vector).    Following Anderson and May (\cite{anderson1992infectious,Favier2005} ) model, we assume fixed incubation period of hosts and 
vectors, constant host recovery rate and vector mortality rate.  Setting fixed incubation periods  allows us to skip the exposed population 
from the dynamics.  The total number of humans $N_H$ is assumed constant, and human mortality is neglected as human life span is much larger than the time scale of infection and recovery. Dengue death rates are rare, and we assume in the current model that the Dengue infection does not progress into serious condition like dengue hemorrhagic fever which could be fatal.  The birth and death rate of 
vectors is assumed constant, leading to constant vector population $N_v$, and direct effect of environmental parameters on vector population 
is not considered \cite{chowell2007estimation}.  The temporal evolution of the \textit{fraction} of susceptible, infected and recovered humans ($s_h, i_h,r_h$), and 
\textit{fraction} of susceptible and infected vectors ($s_v,i_v$) are given by the following delay differential equations 
\cite{chowell2007estimation}:
\begin{eqnarray}\label{eq1}
 \frac{d s_h}{d t} &=& -mC\beta_{hv}  i_v(t) s_h(t) \nonumber \\
 \frac{d i_h}{d t} &=&mC\beta_{hv}  i_v(t-\tau_i) s_h(t-\tau_i)-\gamma_h i_h(t) \nonumber \\
 \frac{d r_h}{d t} &=& \gamma_h i_h(t) \nonumber \\
 \frac{d s_v}{d t}&=& \mu_v n_v(t)  -C\beta_{vh} i_h(t)s_v (t)-\mu_v s_v(t) \nonumber \\
 \frac{d i_v}{d t}&=& e^{-\mu_v \tau_e}C\beta_{vh} i_h(t-\tau_e)s_v (t-\tau_e)  -\mu_v i_v(t) .
\end{eqnarray}
The susceptible humans $s_h$ is infected with rate $mC\beta_{hv}i_v(t)$, and susceptible vectors are infected with 
rate $C\beta_{vh}i_h(t)$. The delay terms account for the incubation period in hosts.  The mean duration of infectious period of  
humans is given by $\frac{1}{\gamma_h}$ and mean adult life span of mosquitoes $\frac{1}{\mu_v}$.  The term $e^{-\mu_v \tau_e}$ accounts 
for the vector mortality during the incubation period.  These differential equations can be evaluated numerically with given initial 
conditions to understand the temporal evolution of the disease in population and eventual steady state. Our focus here is \textit{not} on 
temporal evolution but rather on estimating the rapidity of spread in the initial phase as described below.

\subsection{Reproduction number}

The \textit{basic reproduction number} $R_0$ is used to assess whether the disease propagation reaches an epidemic scale or dies down eventually.
$R_0$ gives the expected number of secondary infections produced from a primary infected individual. For vector borne diseases following the compartmental model in Eq \ref{eq1}, 
nonlinear fixed point analysis gives the reproduction number  \cite{anderson1992infectious,chowell2007estimation}  as: 
 \begin{equation}
 R_0=\frac{mC^2\beta_{hv}\beta_{vh}}{\mu_v\gamma_h}e^{-\mu_v\tau_e}.
 \label{R0_Direct}
\end{equation}

In practice this method is not suitable as certain parameters like transmission rates and mosquitoes per person are not easily measured. However, the intrinsic growth rate  during the initial phase of epidemic carriers useful information about these rates, obviating the need to specially measure these parameters directly. Also, people who are immune to the Dengue due to prior exposure to the infection cannot participate in the disease dynamics. In practice, the reproduction number calculated is $R_p=(1-p)R_0$, which excludes the fraction of immune people ($p$) at the start of the epidemic. Favier \textit{et al} \cite{favier2006early} derived the expression for reproduction number in terms of initial growth rate $\Lambda$, incubation periods and rates (see Appendix for details):
\begin{equation}
R_p=\left(1+\frac{\Lambda}{\gamma_h}\right)\left(1+\frac{\Lambda}{\mu_v}\right)e^{\Lambda (\tau_e+\tau_i)}.
\label{R0_Growth}
\end{equation}
This method requires knowledge of distribution of host parameters $(\tau_e,\tau_i), (\mu_v,\gamma_h)$, which  has been studied extensively.  The mean incubation period in humans is 5.5 days, and in \textit{A. Aegepti} mosquitoes it is  10 days \cite{gubler1998dengue},   and both follow Gamma   distribution. The mean host infection period $1/\gamma_h$ and adult mosquito life span are 5 and 10 days respectively \cite{chowell2013basic}, and also follow Gamma distribution. In table \ref{tab:table1}, we summarize the values and distributions with their references. The growth rate $\Lambda$ is estimated from the time series as below.

\begin{table}[ht!]
\centering
\caption{Model parameters and their corresponding distributions used in the estimation of the reproduction number}
\label{tab:table1}
\begin{tabular}{|l|l|l|}
\hline
\textbf{Parameter}                          & \textbf{Mean(95\% CI)} & \textbf{\begin{tabular}[c]{@{}l@{}}Probability \\ distribution\end{tabular}} \\ \hline
Growth rate ($\Lambda$)                       & Table-2                & t-distribution                                                               \\ \hline
Intrinsic incubation period($\tau_i$)       & 5.5 (4, 7) days \cite{gubler1998dengue}       & Gamma(53.8, 0.1)                                                             \\ \hline
Host infection period($\frac{1}{\gamma_h}$) & { 5.0} (3, 7) days \cite{gubler1998dengue}        & { Gamma(25, 0.2)}                                                              \\ \hline
Adult mosquito life span($\frac{1}{\mu_v}$) & 10 (7, 13) days \cite{chowell2013basic}        & Gamma(44.4, 0.2)                                                             \\ \hline
Extrinsic incubation period ($\tau_e$)      & 10 (8, 12) days      \cite{gubler1998dengue}  & Gamma(100, 0.1)                                                              \\ \hline
\end{tabular}
\end{table}

\begin{figure}[ht!]
\centering 
\fbox{\includegraphics[clip, trim=2.5cm 21cm 3.5cm 2.5cm, scale=0.8]{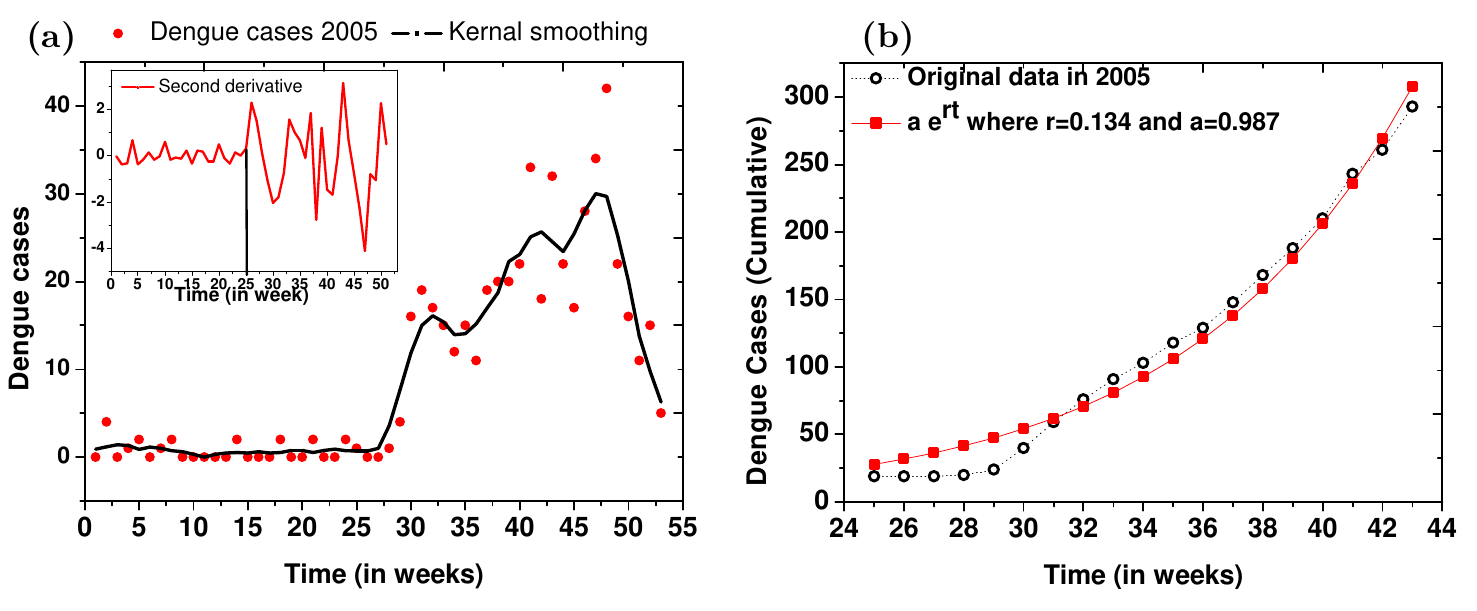}}
\caption{(a) Kernel smoothing of Dengue time series  data (2005) and its second derivative  (b) Exponential fit for cumulative Dengue cases in 2005. }
\label{fig_smoothsecondir}
\end{figure}
 
To calculate the initial growth rate, we observe that there has been outbreak every year roughly around the monsoon season in 
Ahmedabad since 2005. But the data is too noisy to directly decipher the onset of epidemic and its initial growth rate each year. 
We take time series for each year, and smoothen it by convolving it with a Gaussian Kernel. The derivatives of the smoothened data 
is used to estimate the start of the epidemic, which we call $t_0$. We take $n$ data points from $t_0$ where $n$ minimizes the  $R^2$ 
value from an exponential fit, and set the final point to $t_0+n-1$. The growth rate $\Lambda$ is  computed from the exponential fit to 
the original cumulative number of cases in the interval $[t_0,t_0+n-1]$. In Fig~\ref{fig_smoothsecondir}, we illustrate this process for 
the year 2005. Panel (a) gives the kernel smoothened data and its second derivative. In Panel (b), we show the cumulative number of 
Dengue cases from start of the epidemic (25 weeks), and corresponding exponential fit (each of our fits, the $R^2>0.98$). We compute the reproduction number $R_p$ by using $\Lambda$ and parameters described above. The uncertainty in $R_p$ is calculated by sampling the parameters $\tau_e,\tau_i,\gamma_h,\mu_v$ from their respective Gamma distributions and $\Lambda$ from Student's $t$ distribution, and estimating the corresponding variance in $R_p$. We choose $10^5$ samples for our calculations.

\begin{table}[ht!]
\centering
{
\caption{Reproduction number}
\label{tab:table2}
\begin{tabular}{|c|K{2.8cm}|c|c|c|c|}
\hline
\textbf{Year} & \textbf{$t_0$ in week (data length)} & \textbf{$\Lambda$ per Week (95\% CI)} & \textbf{$R_p$} (95\% CI)  & \textbf{$\beta_{hv}$}    
\\ \hline
2005 & 25 (18)       & 0.134 (0.121, 0.146)  & 1.753 (1.748, 1.758)     & 0.210                 
  \\ \hline
2006 & 20 (21)     & 0.115 (0.106, 0.124)  & 1.626 (1.622, 1.630)     & 0.202                
     \\ \hline
2007 & 23 (21)      & 0.060 (0.057, 0.062)  & 1.292 (1.290, 1.294)     & 0.180                  
      \\ \hline
2008 & 26 (20)      & 0.093 (0.087, 0.098)  & 1.485 (1.482, 1.488)     & 0.193                 
      \\ \hline
2009 & 25 (18)     & 0.062 (0.060, 0.064)  & 1.308 (1.306, 1.309)     & 0.181                 
       \\ \hline
2010 & 27 (17)     & 0.098 (0.093, 0.103)  & 1.516 (1.513, 1.519)     & 0.195                 
     \\ \hline
2011 & 21 (23)     & 0.126 (0.119, 0.133)  & 1.704 (1.700, 1.708)     & 0.207                 
         \\ \hline
2012 & 20 (22)      & 0.123 (0.117, 0.129)  & 1.686 (1.682, 1.689)     & 0.206                 
         \\ \hline
\end{tabular}
}
\end{table}

We calculate $R_p$ and its uncertainties for every post monsoon epidemic from 2005-2012.
By comparing Eqs. \ref{R0_Direct} and \ref{R0_Growth}, and assuming $\beta_{hv}=\beta_{vh}$ we get an estimate the 
infection transmission rate $\beta_{hv}$. In table \ref{tab:table2}, we tabulate $t_0$, growth rate $\Lambda$, reproduction 
number $R_p$ and its variance, and estimated transmission rate $\beta_{hv}$ for these years.  We observe that the growth rate 
varies significantly across these years with minimum of 0.060 (2007) to maximum of 0.134 (2005). Correspondingly we see $R_p$ varying  
from 1.29 in 2007 to 1.75  in 2005, and $\beta_{hv}$ between minimum 0.18 to maximum 0.21. The $\beta_{hv}$ is underestimating by a 
factor $1-p$ as we do not have an estimate of the fraction of immune population at the start of epidemic. The year 2005, the growth 
rate was the fastest when probably preventive measures to contain epidemic was not in place. The average value of 
reproduction number $R_p$ and transmission rate ($\beta_{hv}=\beta_{vh}$) from 2005-2012 are 1.54 and 0.197 respectively.
Based on this to bring $R_p<1$, we calculate that mosquito density has to be reduced by atleast 35\%  ($ (R_p-1)/R_p\times 100 $) 
to control the post-monsoon epidemic.

\section{Time series analysis of Dengue cases}

\begin{figure}[ht!]
\centering 
\fbox{\includegraphics[clip, trim=2.0cm 20.5cm 4cm 2.5cm, scale=0.8]{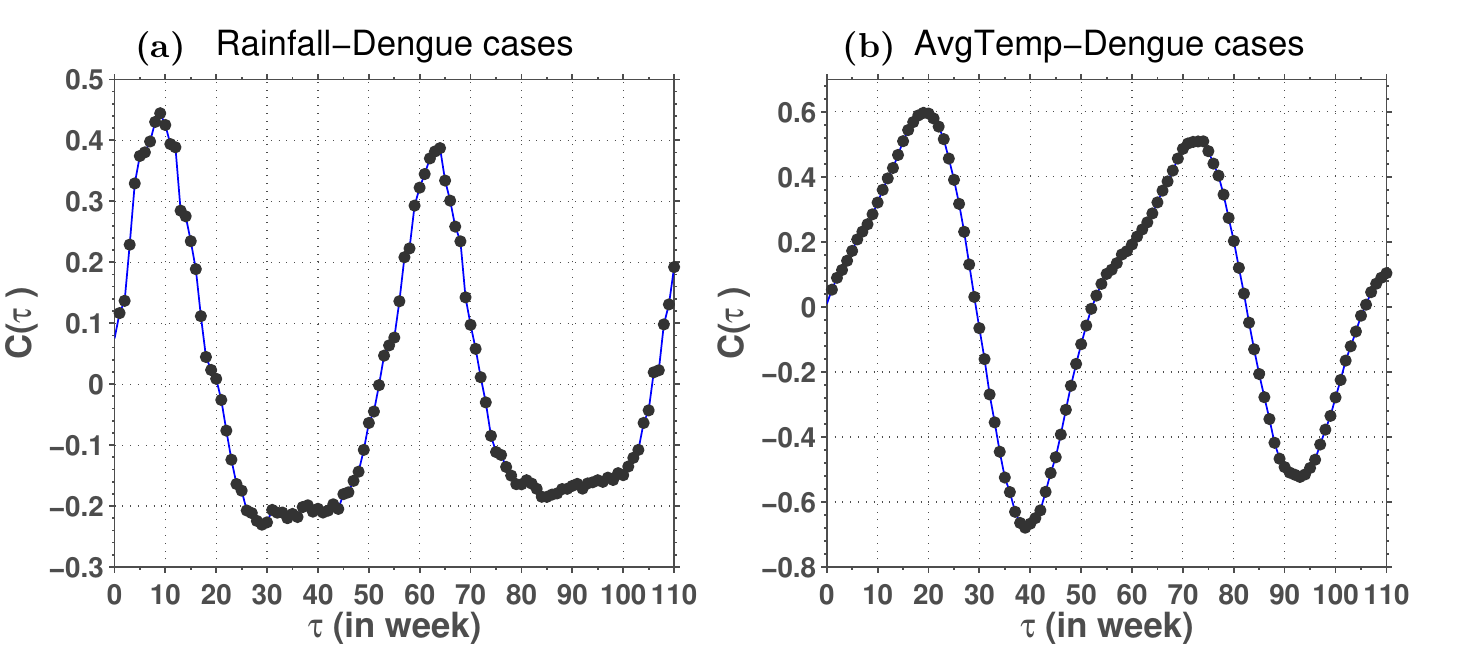}}
\caption{Cross correlation between (a)  Dengue incidences and rainfall (b) Dengue incidences and average temperature}
\label{fig_CRD}
\end{figure}

Compartmental models are useful the disease propagation in a population, when all transmission parameters fixed and initial condition of infection are known. However, it fails to account for the complex dynamics resulting from changes in the environmental conditions, seasonality, human dynamics and other influences. The reproduction number  calculated in previous section gives an estimate of the severity of the epidemic at the threshold of outbreak, but fails to capture the long term trends in the spread of Dengue. Spectral analysis of the time series epidemiological data can reveal the complex temporal dynamics of the disease, including  cycles, and  correlation with environmental parameters like temperature and rainfall \cite{catalano1987time, chatfield1989analysis}. In this work we describe the basic spectral analysis of Dengue with rainfall and average temperature.
Basic comparison of two time series $x(t),y(t)$ can be done through computing cross correlation $C_{XY}(\tau)=(x*y)\tau$ between the signal signals.  In Fig. \ref{fig_CRD}  , we show the cross correlation between Dengue incidences and rainfall (Panel a),  Dengue incidences and average temperature (Panel b). We see from the first peak that there is a lag of 9 weeks between the peak of rainfall and peak of the Dengue incidences. Remaining periodicities are due to the annual cycles of rainfall (monsoon). The temperature and Dengue incidences are anti correlated. At the peak of summer in arid climate of Ahmedabad, Dengue cases are minimal as the virus survival and mosquito breeding  chances are limited. Monsoon follows the peak summer with a gap of  about 10 to 14 weeks, and correspondingly, in Fig. \ref{fig_CRD} panel (b), we a see correlation peaks between temperature and Dengue cases at 20 weeks .  

The classical time series analysis using cross correlation and Fourier spectra completely ignores the association in frequency/periodicities, and is inadequate to analyze  non stationary data \cite{cazelles2005nonstationary, duncan1996whooping}.  Wavelet analysis is one of the effective tools to analyze the non stationary data, through which we can study both time and periodicities (frequency) in one domain. 
 Continuous Wavelet Transform gives spectral components as a function of time of a single data series. Cross comparison of two different time series in time and frequency,  can be done through cross wavelet and wavelet coherence analysis.  Here we give a brief overview of the wavelet analysis for epidemiological time series closely following  \cite{cazelles2007time} and our analysis for Dengue time series.

\subsection{Wavelet Analysis of Dengue and Rainfall time series }
Wavelets consists of a family of basis functions, which can be localized in time and varied in scale. These wavelets $\psi_{a,\tau}$ (called daughter wavelets) are derived from time shifts $\tau$ and scaling $a$ of a single function called the mother wavelet $\psi(t)$ \cite{mallat1999wavelet, daubechies1992ten} as below :
\begin{equation}
\psi_{a,\tau}(t)=\frac{1}{\sqrt{a}}\psi\left( \frac{t-\tau}{a}\right).
\label{Wavelet}
\end{equation}
Here, the scaling parameter $a$ is related to the inverse of frequency.  The mother wavelet integrates to zero $\int_{-\infty}^\infty \psi(t)dt=0$ and is normalized to unity $\int_{-\infty}^\infty |\psi(t)|^2 dt=1$. The most frequently used mother wavelet is the Morlet wavelet, which is a Gaussian with sinusoidal modulation  
\[
\psi(t)=\frac{1}{\pi^{1/4}}\exp(i \omega_0 t)\exp(-t^2/2),
\label{Morlet}
\]
where $\omega_0$ is an arbitrary nonzero number(we choose $\omega_0=3$). The continuous wavelet transform (CWT) of a function $x(t) $ is defined as 
\[
W_x(a,\tau)=\frac{1}{\sqrt{a}}\int_{-\infty}^{\infty} x(t) \psi^*\left( \frac{t-\tau}{a}\right) dt .
\]
The coefficient $W_x(a,\tau)$ gives a measure of the strength of function $x(t)$ at time $\tau$, and scale $a$, and hence allows us to infer the temporal behavior at different times and periodicities. 

The energy density at frequency $\omega$ of signal $x(t)$ is given by the Fourier power spectra $S(\omega)=|F(\omega)|^2$. 
By extending the analogy, the \textit{wavelet power spectrum} is defined as $S(\omega,\tau)=|W_x(\omega,\tau)|^2 $, 
where $\omega=2\pi/a$, gives the energy spectral density at frequency $\omega$ and localized at time $\tau$ \cite{torrence1998practical}. 
The inverse relation between frequency and scale allows us to identify $a$ with the period. 

\begin{figure}[ht!]
\centering 
\fbox{\includegraphics[clip, trim=2.0cm 21cm 3.5cm 2.5cm, scale=0.8]{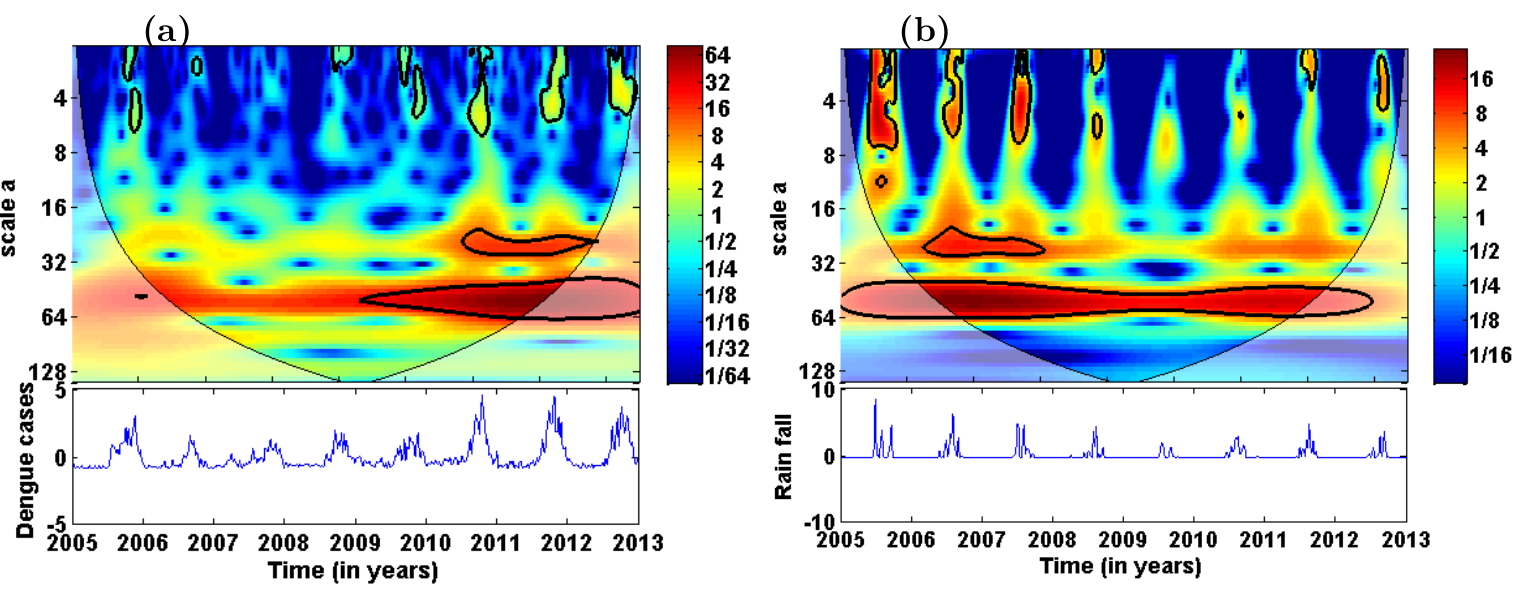}}
\caption{ Wavelet Power spectrum of  (a) Dengue incidences (b)Rain fall }
\label{fig_WPS}
\end{figure}
The wavelet power spectrum of Dengue cases and rain fall are shown in Fig~\ref{fig_WPS}. 
There are many common features between the WPS of these two series such as the significant peaks in the monsoon 
seasons in every year from 2005 to 2012 (see annual longitudinal bands during monsoons for $a$ between 1-8 weeks), and the 
largest band which spreads across the time axis both series around $a=52$ weeks, corresponding to annual periodicity.  The timing and 
duration of the outbreak each year can be seen in the topmost bands in panel a. For example in 2005, the rise of epidemic started 
around 25 weeks from Jan ($t$ between 2005-2006), and lasted for about 10 weeks ($a \in [1,10]$).  The start of the outbreak 
(see Table~\ref{tab:table2}) and scale varies every year. The outbreaks peaked during 2010-2012 lasting 3-4 months. The second largest
band at scale of 30 weeks, is due to overlap of end of the previous year outbreak with the beginning of the outbreak in 
the current year. Rainfall patterns are more predictable show sharp timing and periodicities.  The duration of monsoon was 
longer in the years 2005-2006, to 2007-2008. Duration of monsoon  lasts for 2-3 months. The monsoon peaked in 2005, and it also 
corresponded with a high reproduction number (1.753 see in Table~\ref{tab:table2}).  A closer examination of these relationships 
needs  Wavelet Coherence, discussed in next section.

\subsection{Coherence between Dengue cases and climatic variables}
Wavelet Cross Spectrum and Wavelet Coherence measures are used to capture the statistical relationships between two non-stationary signals. The wavelet cross spectrum, also called Cross Wavelet Transform (XWT) \cite{torrence1999interdecadal} of $x(t),y(t)$ is given by $W_{x,y}(a,\tau)=W_{x}(a,\tau)\cdot W_{y}^*(a,\tau)$. where `*' denotes the complex conjugate.  XWT indicates linear overlap between the two signals (un-normalized) in time and frequency domain. The wavelet coherence $C_{x,y}(a,\tau)$ is the cross-spectrum smoothened over time and scale (expectation value) and normalized by the smoothened spectrum of each time series \cite{torrence1999interdecadal, cazelles2007time}. It allows to explain the causality and coherence between the signals. 
\begin{equation}
C_{x,y}(a,\tau)=\frac{||\langle W_{x,y}(a,\tau) \rangle||}{\sqrt{||\langle W_{x,x}(a,\tau) \rangle||~||\langle W_{y,y}(a,\tau) \rangle||}  }
\end{equation}
$\langle W_{x,y}(a,\tau) \rangle=\int_{a-\Delta/2}^{a+\Delta/2}\int_{\tau-\delta/2}^{\tau+\delta/2} W_{x,y}(\alpha,t)d\alpha U_{\Delta,\delta}(\alpha,t) dt $,
where the weight function $U$ satisfies \\
$ \int_{a-\Delta/2}^{a+\Delta/2}\int_{\tau-\delta/2}^{\tau+\delta/2} U_{\Delta,\delta}(\alpha,t) dt=1 $.
The relative phase between the signals is given by $\displaystyle \phi(a,\tau)=\tan^{-1}\left( \frac{\Im \langle W_{x,y}(a,\tau) \rangle}{\Re \langle W_{x,y}(a,\tau)\rangle} \right  )$. The $C_{x,y}(a,\tau)$ varies between 0 and 1, reaching extreme 1 when there is perfect correlation at a particular time and scale, and zero when the two time series are independent. A constant or uni-modal distribution of relative phase in a particular time-frequency band indicates that time series are phase locked in that band. In the absence of it, distribution is random. 
\begin{figure}[ht!]
\centering 
\fbox{\includegraphics[clip, trim=2.0cm 21.5cm 3.5cm 2.5cm, scale=0.8]{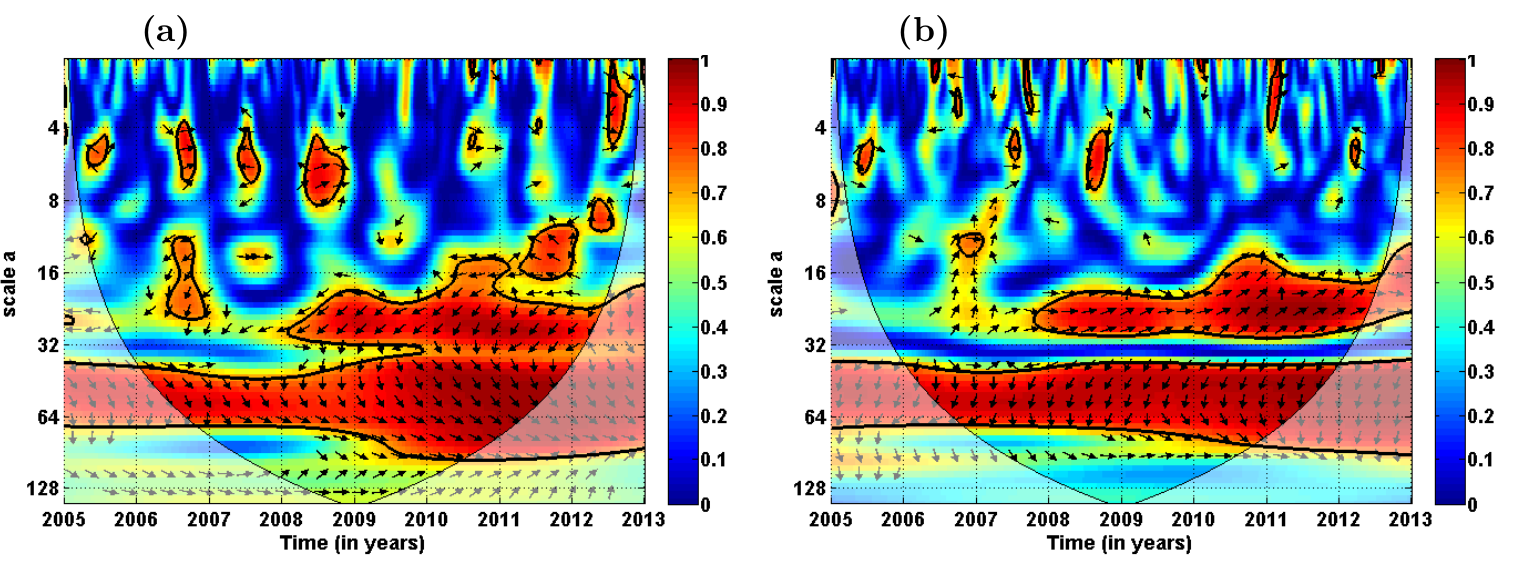}}
\caption{Wavelet coherence between (a) Dengue incidences and rain fall
(b) Dengue incidences and average temperature.}
\label{fig_wtcrfdc}
\end{figure}

In Fig~\ref{fig_wtcrfdc}(a), we show the wavelet coherence of Dengue incidences and 
rain fall. The largest band around the 52 weeks indicates the annual cycles in which both series are phase locked, with Dengue cases lagging behind the rain fall. The second largest band is in the 16-32 weeks band during 2008-2012. Here the phase lock is between 20-30 weeks , indicating a stronger correlation for long duration (2.5-4.5 months).  Wavelet power stronger in the 4-8 weeks  band from 2005-2008, however there seems to be no phase lock in this time period. In 2006-2007, 10-30 weeks band, there is high wavelet correlations and phase locking (7.5 weeks - 22.5 weeks).  As we expected some of the regions (pre monsoon seasons) with low coherence incidences because of the low wavelet power of rain fall.

Vector incubation period and vector mortality rate depend on temperature, and hence influences Dengue incidences. Wavelet coherence between Dengue incidences and average Temperature is shown in Fig~\ref{fig_wtcrfdc}(b). The largest band corresponding to annual periodicity (52 weeks) shows a phase of lag of about -90 degrees. Second largest band lies between 16-32 weeks period from 2008 to mid of 2012. Wavelet power in this band is similar to the coherence spectra of Dengue incidences with rainfall, however the phase angle is widely different. From 2008-mid of 2010, the relative phase is zero implying the that temperature and Dengue incidences are in phase.  Phase difference becomes slightly positive post 2010, in contrast to Dengue-Rainfall. There is a narrow band around 2007 and scale 8-32 weeks, where  the phase lock is about 90 degrees, and  $C_{x,y}$ is about 0.6. In other regions, we do not see much correlations.


\section{Summary and Conclusions}
In this work, we analyzed the Dengue incidence and its correlation with climatic variables in Ahmedabad city from 2005-2012. Using the initial growth rate of Dengue incidences and biological parameters, we were able to estimate the reproduction number $R_p$ for post monsoon outbreaks every year from 2005-2012, which lied between 1.29 to 1.75 . The uncertainties in $R_p$ were estimated using parameter uncertainties and Monte-Carlo methods. We could also estimate the effective transmission rate $\beta_{hv}$ from $R_p$ and compartmental model expression for basic reproduction number $R_0$. Based on these, the minimal reduction in mosquito density required to prevent Dengue outbreak in Ahmedabad has been calculated.  

The long term trends in Dengue series and its climatic correlates were analyzed using Wavelet based spectral methods.  Classical spectral methods were used to estimate the mean time lag between  Dengue incidences and Rainfall (9 weeks), and also between Dengue cases and average temperature. Continuous Wavelet Transform analysis of Dengue cases and rainfall revealed non stationary patterns of the Dengue spread such as variations in start and duration of the epidemic. Wavelet coherence analysis revealed phase locking between Dengue cases and rainfall in the 16-32 week band, even though individual wavelet transforms showed opposing tendencies of increase in duration of outbreak for Dengue, while the monsoon duration decreased over time. Temperature Dengue incidences also show correlations, but no significant phase lock over long periods. 

Our work is based on empirical analysis of time series data and we could analyze the non stationary patterns of Dengue, and correlation with environmental variables at different times and scales. But, a clear causal link between the environmental variables and Dengue incidences, if it exists, in Ahmedabad could not be established in the present study. This demands a closer study of the factors apart from Climatic variables influencing the hyperepidemicity of Dengue such as viral adaptation to na\"ive hosts, socio-economic factors like urbanization and human mobility.

\section*{Acknowledgments} 
The authors would like to thank RamRup Sarkar for insightful comments and Dr. V K Kohli, Assistant Entomologist Ahmedabad Municipal Corporation for providing us with data on Dengue incidences and mosquitoes in Ahmedabad city. We gratefully acknowledge the partial funding for this work from SERB-DST India through project number SB/FTP/PS-033/2013.


\bibliographystyle{AIMS}
\bibliography{references}

\section{APPENDIX}
\subsection{Reproduction number from the intrinsic growth rate}

A brief derivation of the reproduction number in Eq. \ref{R0_Growth} based on
\cite{favier2006early} is given below. We start with the compartmental equations Eq.\ref{eq1} for infected humans and vectors
\begin{eqnarray}\label{InfHumVec}
 \frac{d i_h}{d t} &=&mC\beta_{hv}  i_v(t-\tau_i) s_h(t-\tau_i)-\gamma_h i_h(t) \nonumber \\
 \frac{d i_v}{d t}&=& e^{-\mu_v \tau_e}C\beta_{vh} i_h(t-\tau_e)s_v (t-\tau_e)  -\mu_v i_v(t) .
 \end{eqnarray}

At the onset of epidemic, the exponential growth of fraction of infected hosts and vectors is given by $i_h\simeq i_{h0}e^{\Lambda t}, i_v \simeq i_{v0}e^{\Lambda t}$, $s_h\simeq 1, s_v \simeq 1 $.  Plugging this in Eq. \ref{InfHumVec}, we get:

\begin{eqnarray}
i_{h0}(\Lambda +\gamma_h ) e^{\Lambda t} &= &mC\beta_{hv} i_{v0} e^{\Lambda (t-\tau_i)} \nonumber \\
i_{v0}(\Lambda +\mu_v ) e^{\Lambda t} &= & e^{-\mu_v \tau_e}C\beta_{vh} i_{h0} e^{\Lambda (t-\tau_e)} 
\end{eqnarray}
Multiplying  both the equations and with basic algebra, we get.
\begin{equation}
R_0=\frac{mC^2\beta_{hv}\beta_{vh} e^{-\mu_v\tau_e}}{\mu_v\gamma_h}=\left(1+\frac{\Lambda}{\mu_v}\right)\left(1+\frac{\Lambda}{\gamma_h}\right)e^{\Lambda (\tau_e+\tau_i)}
\end{equation}

\subsection{Calculation of $R_p$ and distribution of parameters}

\begin{figure}[ht!]
\centering
\includegraphics[clip, trim=2.5cm 3cm 3.4cm 3cm,scale=0.8]{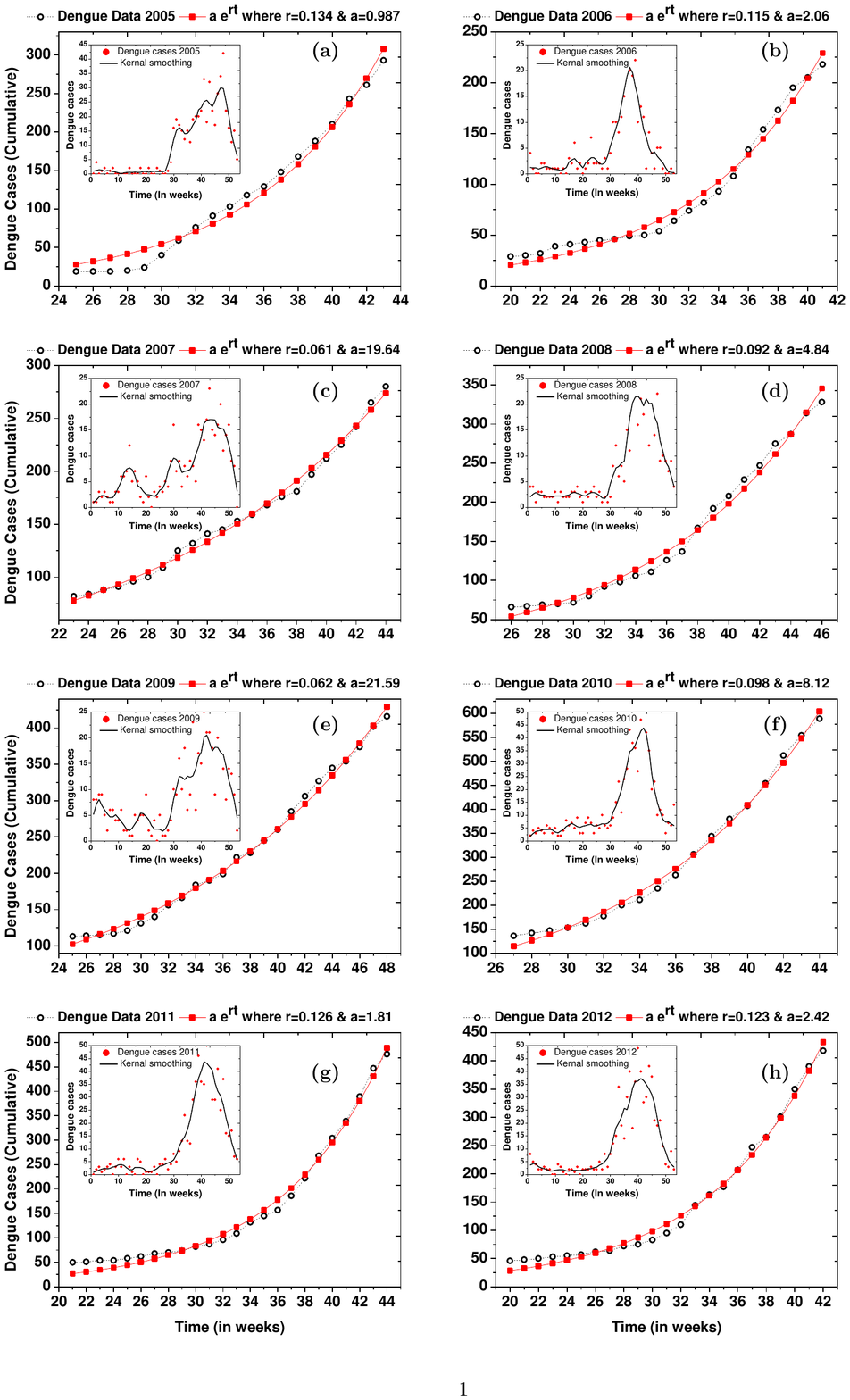}
\caption{Exponential fit for the cumulative Dengue cases for the years 2005-2012. Inset: Their corresponding Kernel smoothing of Dengue time series data}
\label{ExpFitAll}
\end{figure}

In Figure \ref{ExpFitAll}, we give the Kernel smoothing of Dengue time series data, exponential fit for the cumulative Dengue cases post monsoon epidemic outbreak for the years 2005-2012, extension of Fig 2 in main text.
 
\begin{figure}[h!]
\centering
\includegraphics[clip, trim=2cm 16cm 3.4cm 3cm,scale=0.8]{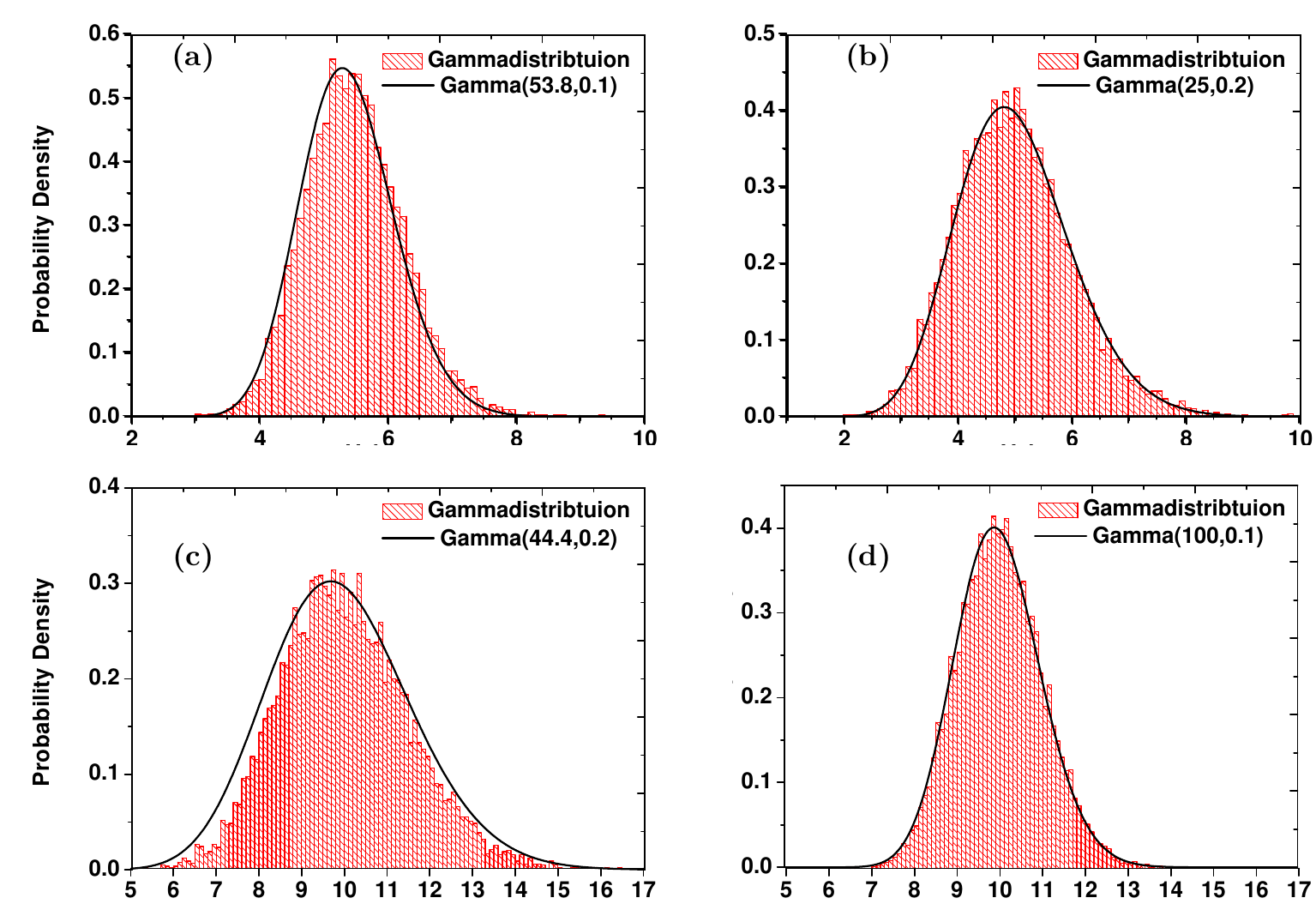}
\caption{Distribution of model parameters discussed in Table 1 (a) Intrinsic incubation period $\tau_i$ (b) Host infection period $1/\gamma_h$ (c) Adult mosquito life span $1/\mu_v$ (d) Extrinsic incubation period $1/\tau_e$}
\label{RPforyearAll}
\end{figure}

In Figure \ref{RPforyearAll}, we give the Distribution of model parameters in Table I, used in the estimation of the reproduction number. 

\end{document}